\documentclass[12pt,preprint]{aastex}

\def\kms{$\rm km\;s^{-1}$} 
\def\kmsmpc{$\rm km\;s^{-1}\;Mpc^{-1}$}

\shortauthors{Pizzella et al.} 
\shorttitle{Nuclear Stellar Disks in Spiral Galaxies} 
 
\begin{document} 
 
\title{Nuclear Stellar Disks in Spiral Galaxies\thanks{
Based on observations with the
NASA/ESA {\it Hubble Space Telescope\/},
obtained from the data Archive at the Space
Telescope Science Institute,
which is operated by the Association
of the Universities for Research in Astronomy,
Inc., under NASA contract NAS 5-26555. 
}} 
 
\author{
A.~Pizzella\altaffilmark{1},
E.M.~Corsini\altaffilmark{1}, 
L.~Morelli\altaffilmark{1},
M.~Sarzi\altaffilmark{1,2}, 
C.~Scarlata\altaffilmark{1,3},
M.~Stiavelli\altaffilmark{3} and
F.~Bertola\altaffilmark{1}}
\altaffiltext{1}{Dipartimento di Astronomia, Universit\`a di Padova, 
Vicolo dell'Osservatorio 2, I-35122 Padova, Italy} 
\altaffiltext{2}{Max-Planck-Institut f\"ur Astronomie, Koenigstuhl 17,  
D-69117 Heidelberg, Germany} 
\altaffiltext{3}{Space Telescope Science Institute, 3700 San Martin 
Dr., Baltimore, MD 21218, USA}

\begin{abstract} 

We report evidence for nuclear stellar disks in 3
early-type spirals, namely NGC 1425, NGC 3898 and NGC 4698,
revealed by  WFPC2/F606W images out of a sample of 38 
spiral galaxies, selected from the 
{\it Hubble Space Telescope\/} Data Archive. Adopting 
the photometric method introduced by Scorza
\& Bender we derived their central surface brightness and
scalelength by assuming them to be infinitesimally thin exponential
disks. No nuclear disk was found in barred galaxies or galaxies of 
Hubble type later than Sb.
The external origin of the disk in NGC 4698 is strongly suggested by its
orthogonal geometrical decoupling with respect to the host galaxy.

\end{abstract}

 
%
\section{Introduction} 
 
In the last decade, embedded stellar disks or disky distorsions
have been found in many elliptical and S0
galaxies (Scorza \& Bender 1995 hereafter SB95; Seifert
\& Scorza 1996). They are characterized by a 
smaller scalelength and higher central
surface brightness with respect to the large kpc-scale disks typical
of lenticular and spiral galaxies.
The existence of embedded disks gives further observational support
to the idea that the disky ellipticals are the continuation of the 
sequence ranging from Im's through spirals to S0's
(Kormendy \& Bender 1996 and references therein).
This suggests also a continuity in the formation history, whereby
one or several parameters of the protogalaxy vary smoothly (e.g. van den Bosch 1998).

Subarcsecond resolution provided by state-of-the-art ground
and space-based telescopes  have revealed
even smaller stellar disks in the nuclear region of nearby galaxies.
To date the smallest disks we know have scalelengths of few
tens of pc and they have been identified in the nucleus of a handful of
early-type disk galaxies: the S0's NGC~3115 (SB95;
Lauer et al. 1995), NGC~4342 (van den Bosch, Jaffe, \& van der Marel 1998; Scorza \& van
den Bosch 1998), and S0 NGC~4570 (van den Bosch et al. 1998; Scorza \&
van den Bosch 1998; van den Bosch \& Emsellem 1998), and the Sa NGC
4594 (Burkhead 1986, 1991; Kormendy 1988; Emsellem et al. 1994, 1996).
In addition, nuclear stellar disks of gas, dust and stars have been found
in the E3 galaxy NGC~5845 (Kormendy et al. 1994) and in the dwarf E2 galaxy NGC 4486A
(Kormendy et al. 2002). Although the photometric parameters are not known,
these stellar disks appear to be similar to those of NGC 3115 and NGC 4594.
This phenomenon may be more common if some of the E and S0 galaxies with 
disky isophotes observed by Ravindranath et al. (2001) and Rest et al. (2001) turn out to
host a nuclear stellar disk.

The presence of nuclear disks raises the question about the epoch
(i.e. coeval or not with that of the host) and mechanism
(i.e. external or internal origin) of their formation. 
Indeed, it is possible that they formed with a different 
mechanism than the one that formed normal S0 disks since their size
is one order of magnitude smaller.
The blue colour
of the nuclear disks of NGC~4342, NGC~4570 and NGC~4486A suggests they are made
of younger stars with respect to the bulk of their host galaxy (van
den Bosch et al. 1998; Kormendy et al. 2002). In the framework of galaxy formation via
hierarchical merging acquired gas may end up forming some 
of these nuclear stellar disks. On the other
hand, they could be built up from gas transported towards the galaxy
center during the secular evolution of a bar. This seems to be the
case of NGC 4570 (van den Bosch \& Emsellem 1998), where the observed 
features in photometry and kinematics correspond with the 
position of the main resonances of a
small bar.
Furthermore, any model for the formation of nuclear disks has to
account for supermassive black holes,
which are supposed to reside in every galaxy and whose gravitational
influence may extend to pc scale of the disks. Galaxies hosting
nuclear disks are not exceptions in this context as NGC 3115, NGC 4342,
NGC 4594, and NGC~5845 are known to harbour a central 
supermassive black hole (Kormendy et al. 1996a,b; Cretton \& van den Bosch 1999, 
Kormendy \& Gebhardt 2002).

To date studies on nuclear stellar disks considered mainly elliptical
or S0 galaxies but there is no a priori reason why they should not be present 
also in spiral galaxies. Would this be true, such nuclear disks would be the result of
star formation by gas either driven toward the center by a bar instability or acquired 
from the outside.

For these reasons we undertook a search for nuclear stellar
disks in spiral galaxies based on HST/WFPC2 archive images. In this
paper we present as a result of our investigation three new cases of
spiral galaxies hosting a nuclear disk embedded in their main outer
disk and bulge.

%
\section{Searching for nuclear stellar disks in spirals} 
 
\subsection{Sample selection} 
 
We selected RC3 (de Vaucouleurs et al. 1991)
galaxies classified as spiral ($T\,\ge\,0$),
and with $cz\,<\,2000$ \kms . 
We restricted our sample to galaxies within $20$ Mpc (for $H_0=100$ \kmsmpc)
to be able to detect 10-pc scalelength nuclear disks on WFPC2/PC frames out to
about $2\,$--$\,3$ scalelengths. We did not exclude low-inclined
galaxies from the sample since nuclear disks might have a different
inclination with respect to their hosts. 
We searched the HST science archive looking for all the available
images. We realized that WFPC2/F606W was the only camera/filter
combination producing an homogeneous sample large enough for our purposes (159
galaxies). This resulted in a sample of 112 objects after we rejected
those frames where galaxy nucleus was out or too close to the chip
edge as well as those with too short exposure times.

It has to be noted that all but 3 
of the selected galaxies belong to 4
SNAP observing programs: (1) 57 objects were taken from the sample of
disk galaxies ranging from E/S0 to spirals observed in Program ID 5446
(PI: G. Illingworth); (2) 33 are unbarred Sa/Sb spirals from Program
ID 6359 (PI: M. Stiavelli); (3) 14 spirals were taken from the nearby
Seyfert galaxies observed in Program ID 5479 (PI: M. Malkan); (4) 5
are Seyfert spirals from Program ID 8597 (PI: M. Regan).
In our final sample the Hubble types ranging from Sa to Scd are well
represented as well as the number of barred spirals roughly equals
that of unbarred ones, although images were collected from
HST programs based on different selection criteria.

\subsection{Analysis of the HST archive images} 
 
The on-the-fly calibrated WFPC2/F606W images of the 112 sample galaxies
were retrieved from the HST archive. The different images of the same
target were aligned and combined.  Cosmic ray events were removed with
IRAF tasks CRREJ and IMEDIT.  The conversion to the Johnson system has
been calculated using SYNPHOT in STSDAS. Since this correction depends
on the spectral energy distribution of the object, it has been calculated using the Kinney
et al. (1996) spectral templates.

As a first step to identify candidates hosting a nuclear disk we
construct unsharp-masked images for all the sample galaxies, dividing
each WFPC2/F606W frame by itself after being convolved with a circular
Gaussian of $\sigma=$ 2,4, and 6 pixels, respectively (Fig. 1). The advantage
of this procedure is to quickly enhance any surface-brightness
fluctuation and non-circular structure extending over a spatial region
comparable with the $\sigma$ of the smoothing Gaussian.

In the present context this enables first to set apart 74 galaxies whose nucleus
is strongly affected by dusts, preventing any further analysis. This is the case
of NGC~1637 shown, as an example, in Fig. 1. It should be noted that, in principle,
one can check for the presence of dust lanes using images with different filters
which are not available for the whole galaxy sample. However adopting different values 
for $\sigma$ in the unsharp-mask allows to clearly reveal dust 
lanes, when present.

Successively, on the unsharp-masked images of the 38 remaining galaxies,
we looked for the highly flattened
nuclear structures  which are possibly inclined nuclear disks.
Such flattened structures are not artifacts of the unsharp-masking
procedure since they are always associated to a central increase of
ellipticity as measured by performing the isophotal analysis using the
IRAF task ELLIPSE. 
This subset of 38 galaxies ranges from Sa to Sm and from $M_{B_T}^0 = -15.6$ to $-20.0$
as shown in Fig. 2 and represents
our qualified sample for which we are confident we are able to detect an inclined
nuclear disk, if present.
We found three objects, namely NGC 1425, NGC 3898 and NGC 4698, which
show this flattened nuclear structure (Fig. 1).
A positive fourth cosine Fourier coefficient (describing the disky
deviation of the isophotes from pure ellipses) confirms the presence
of a nuclear stellar disk in inner regions of these early-type
spirals (Fig. 3).
In Fig. 1 we show the case of NGC~4539 as an example of those galaxies
where we did not identify any nuclear flattened structure.

\subsection{Photometric decomposition} 
 
Once the existence of the nuclear disks is established, we derived their 
photometrical properties by using the method described by SB95.
When adopting this technique to study the innermost regions of
galaxies it is essential to restore the images from the effects of the
HST point spread function (PSF) in order to properly derive the
nuclear disks parameters, as shown by Scorza \& van den Bosch (1998).
Such deconvolution was performed through the Richardson-Lucy method
by means of the IRAF task LUCY. Although susceptible to noise
amplification, this algorithm has been proved by van den Bosch et al.
(1998) to lead to a restored surface-brightness distribution
comparable to the one obtained by means of a multi-Gaussian
representation (Monnet, Bacon, \& Emsellem 1992). We decided
to deconvolve the images with a number of iterations between 3 and 6.
A larger number of iterations does not affect the result of the
decomposition but amplify the noise.  We believe that the
results obtained by van den Bosch et al. (1998) are directly
applicable to our case: we are dealing with images obtained
with similar or longer integration times, galaxies with less steep
surface-brightness profiles and nuclear disks with equal or larger
scalelengths. 

For each given image and nucleus position on the PC (NGC 3898, NGC
4698) or WF2 CCD (NGC 1425) we adopted a model PSF calculated using
the TINY-TIM package (Krist \& Hook 1999).
No correction for telescope jitter was necessary.
The SB95 method consists in the iterative subtraction from
the galaxy image of a thin disk model. The parameters of such a disk
are varied until the departures from perfect ellipses are smallest
(i.e. $a_{4}$ and $a_{6}$ are nearly zero). For the disk component we
assumed an exponential surface brightness profile, with central
surface brightness $\mu_0$, radial scalelength $h$, and an
inclination given by $i=\arccos{(b/a)}$.
We verified that the parameters of nuclear disks resulting from   
the photometric decomposition are not affected by 
small changes in PSF (e.g. 
its generation in different chip positions).

%
\section{Results} 
 
The results of the photometric decomposition of the surface-brightness
distribution of NGC 1425, NGC 3898 and NGC 4698 are shown in Fig.~3
and Fig.~4. In Fig. 3 we plot the ellipticity, position angle, $a_4$, and
$a_6$ Fourier coefficient radial profiles before and after the nuclear
disk subtraction. In Fig. 4 the same is true for the galaxy isophotes.
The photometric parameters derived for the nuclear disks are given
in Tab. 1. Here we discuss the individual objects.

\noindent 
{\bf NGC 1425.} To obtain a residual bulge with nearly elliptical
isophotes we needed to subtract two exponential thin disks of different
scalelengths, central surface brightnesses and inclinations. If this
difference in inclination is real we are facing two structures in an
unstable configuration.

As discussed by Scorza \& van den Bosch (1998) for the the nuclear and
main disks of NGC 4342, it is more likely we are looking at two thick
disks of different thickness (e.g., the exponential spheroid disks of
van den Bosch \& de Zeeuw 1996) rather than two infinitesimally thin
disks of different inclination. 
Such a different thickness may be interpreted as due to the presence
of two distinct exponential disks. Alternatively we may think we are
facing a non-exponential disk with tickness varying with radius.
Considering the two disk hypotesis, we note that the photometric properties 
of the larger-scale disk are close to those observed for disky ellipticals (Fig. 5).
 
\noindent 
{\bf NGC 3898.} The galaxy hosts a small ($h\,=\,18$ pc) nuclear disk
that is noticeable in particular from the Fourier
$a_4$ coefficient positive values, observed within $0\farcs6$. In terms of 
physical size this is the smallest disk so far detected.
It possibly corresponds to the somewhat steeper rise of the stellar
rotation curve measured in the inner $1''$ and the two relative maxima
in the stellar velocity dispersion at $\pm 1''$ by Vega Beltr\'an et al.
(2001) but not considered in the dynamical modelling of Pignatelli et
al. (2001).
 
\noindent 
{\bf NGC 4698.}  Although Sandage \& Bedke (1994) include this galaxy
in The Carnegie Atlas of Galaxies as an example of the
early-to-intermediate Sa type, NGC 4698 has photometrical and
kinematical properties which are uncommon among spiral galaxies. It
shows a clear geometrical decoupling between bulge and disk and hosts
a kinematically isolated core (Bertola et al. 1999).
The nuclear stellar disk corresponds to this isolated core, which is
rotating perpendicularly with respect to the galaxy main disk (Bertola et
al. 1999).

%
\section{Discussion and conclusions} 
  
We have provided evidence for the presence of a nuclear disk in 3 early-type spirals, namely
NGC 1425, NGC 3898 and NGC 4698 over a qualified sample
of 38 objects. 
The photometric properties of these 20-pc scale exponential disks are
consistent with those of the 4 nuclear disks so far detected in disk
galaxies of even earlier morphological type (Fig. 5). We have not
found nuclear disks neither in Sbc--Sm nor in barred galaxies
although these classes represents the majority of our sample.
To further address the demography of nuclear disks we need to apply
the relatively easy approach adopted here (based on unsharp-masking) on
a larger sample of galaxies, imaged at high spatial resolution in
near-infrared pass-bands to deal with central dusts which prevented us
to exclude the presence of nuclear disks in a large fraction (roughly two thirds)
of our WFPC2/F606W sample galaxies.

Without drawing general conclusions from such a limited
number of galaxies, nevertheless we have the indication that the presence of nuclear
disks is restricted to S0's and bulge-dominated unbarred spiral
galaxies. In the framework of massive bulge formation through a
process of hierarchical clustering merging, the nuclear disks may be
the final result of dissipational and star-formation processes
subsequent to a second acquisition event. 
To date however, there were only photometric and kinematical
evidences that nuclear disks can also be formed via secular
evolution of a bar (e.g. NGC 4570).

With NGC 4698 we showed for the first time that second events
indeed represent a viable mechanism to build a nuclear disk in the
center of disk galaxies. Indeed the nuclear disk of NGC 4698 is
geometrically (this paper) and
kinematically (Bertola et al. 1999) decoupled in an orthogonal way
with respect to the host galaxy. This phenomenon can be hardly
explained without invoking the acquisition of external material from
the galaxy outskirts (see Bertola \& Corsini 2000).

\acknowledgments 
We thank F. van den Bosch for providing us his data for Fig. 5.

\begin{figure} 
\epsscale{0.65} 
\plotone{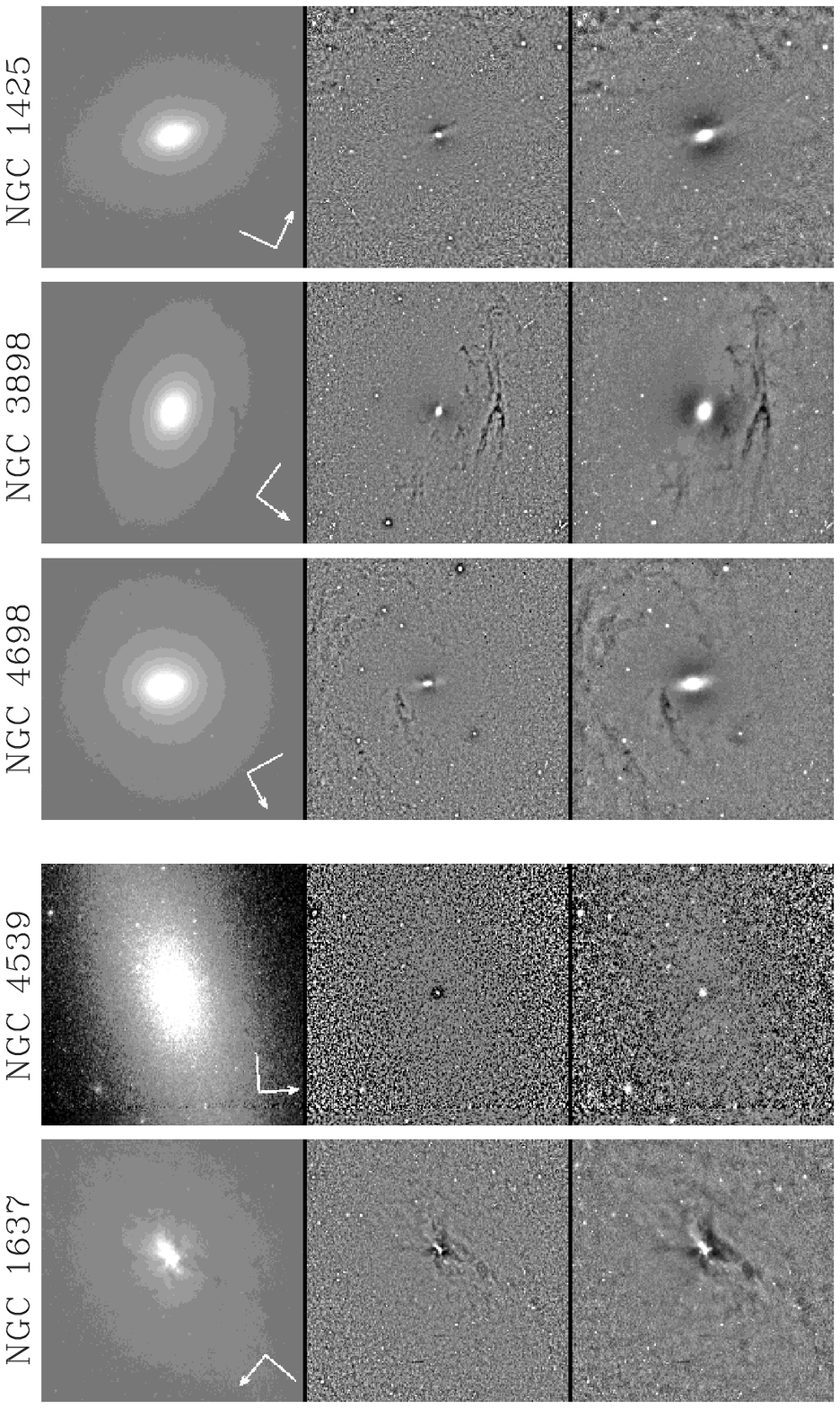} 
\caption{ {\it Left panels:\/} WFPC2/F606W images of NGC 1425, NGC
  3898 and NGC 4698 (where we found a nuclear disk), NGC 4539  (where
  a nuclear disk is not present), and NGC 1637 (where
  dust patches prevent any further analysis).  The size of the plotted region is
  $19\farcs3\times19\farcs3$. The orientation is specified by the arrow
  indicating north and the segment indicating east in the lower right corner
  of each panel. {\it Middle and right panels:\/} Unsharp masking of the
  WFPC2/F606W images obtained with $\sigma=2$ and $6$ pixels,
  respectively. Sizes and orientations are the same as in the left
  panels.}
\end{figure} 
\clearpage

\begin{figure} 
\epsscale{1.} 
\plotone{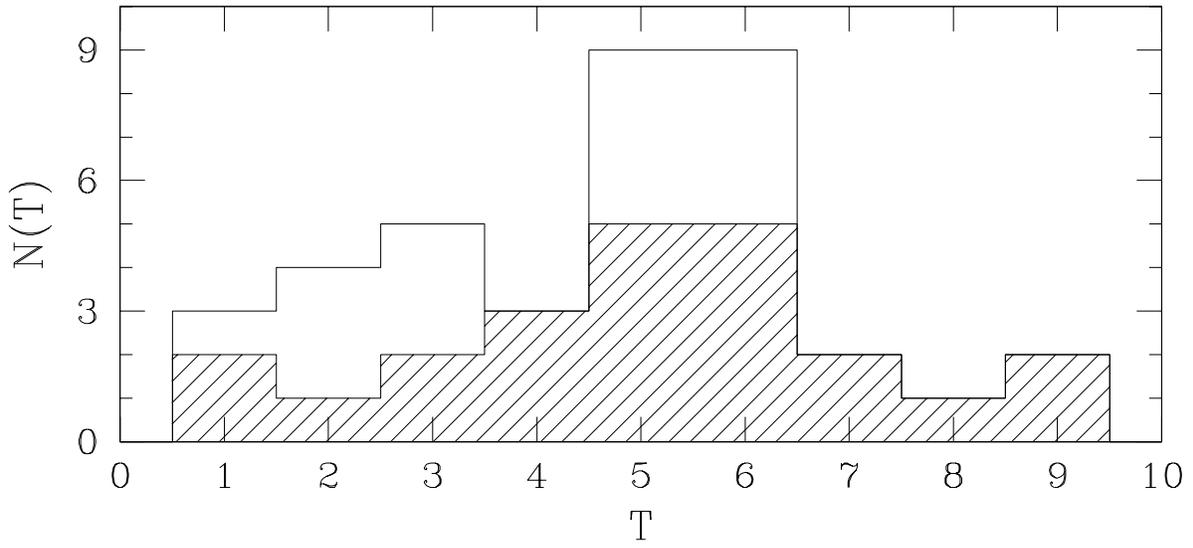} 
\figcaption{Hubble type distribution for the qualified sample of 38 
spiral galaxies.
The dashed region identifies galaxies with
and SB or SAB classification.
 }
\end{figure} 
\clearpage
\begin{figure} 
\epsscale{0.9}  
\plotone{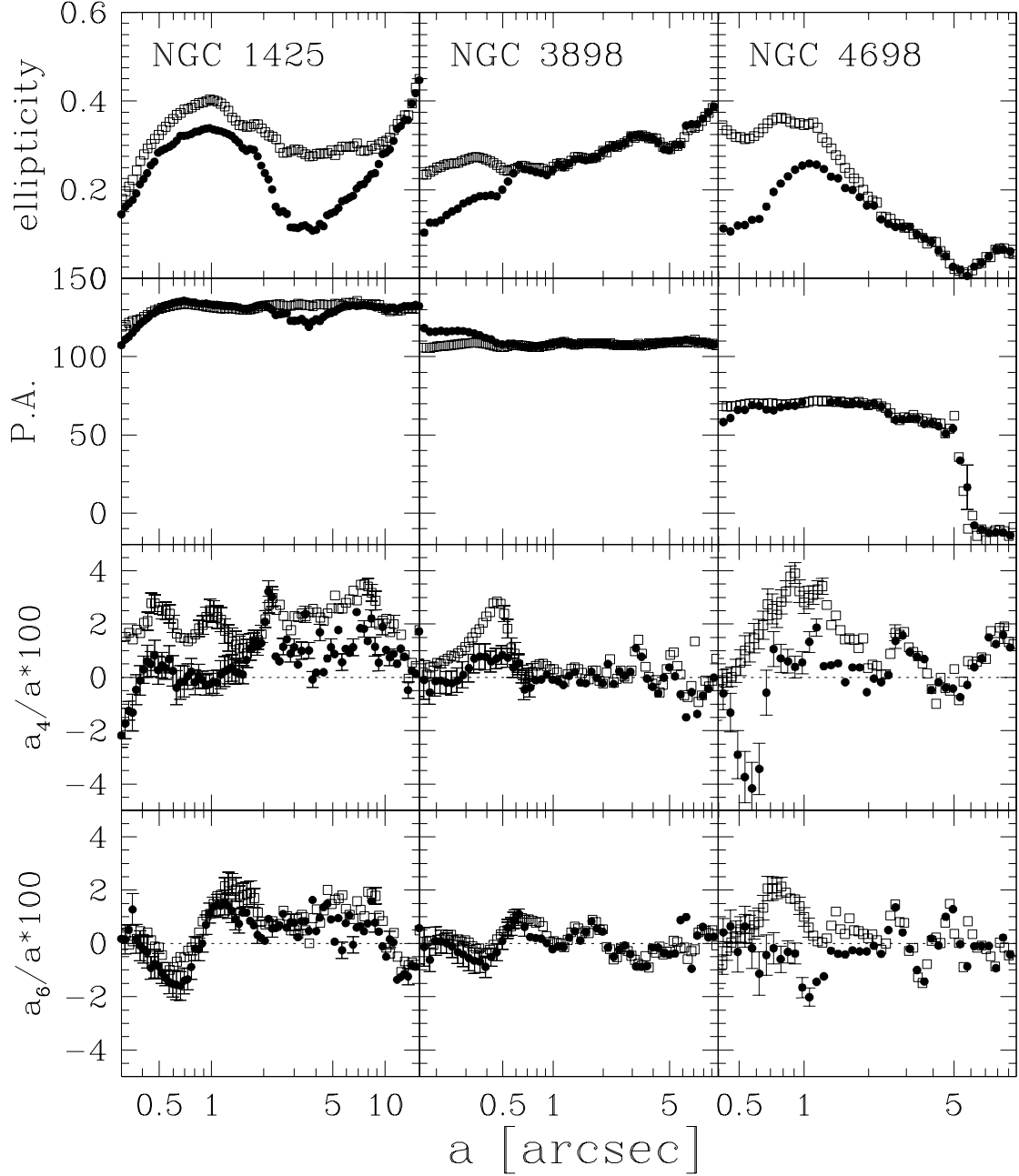}  
\figcaption{Deconvolved radial profiles of $\epsilon$, PA, $a_4$ and
  $a_6$ Fourier coefficients measured before ({\it open squares\/}) and
  after ({\it filled circles\/}) the subtraction of nuclear disks of
  Tab. 1 for NGC 1425 ({\it left panels\/}), NGC 3898 ({\it middle
  panels\/}) and NGC 4698 ({\it right panels\/}). Error bars smaller than
symbols are not plotted.}
\end{figure}  
\clearpage

\begin{figure} 
\epsscale{0.75} 
\plotone{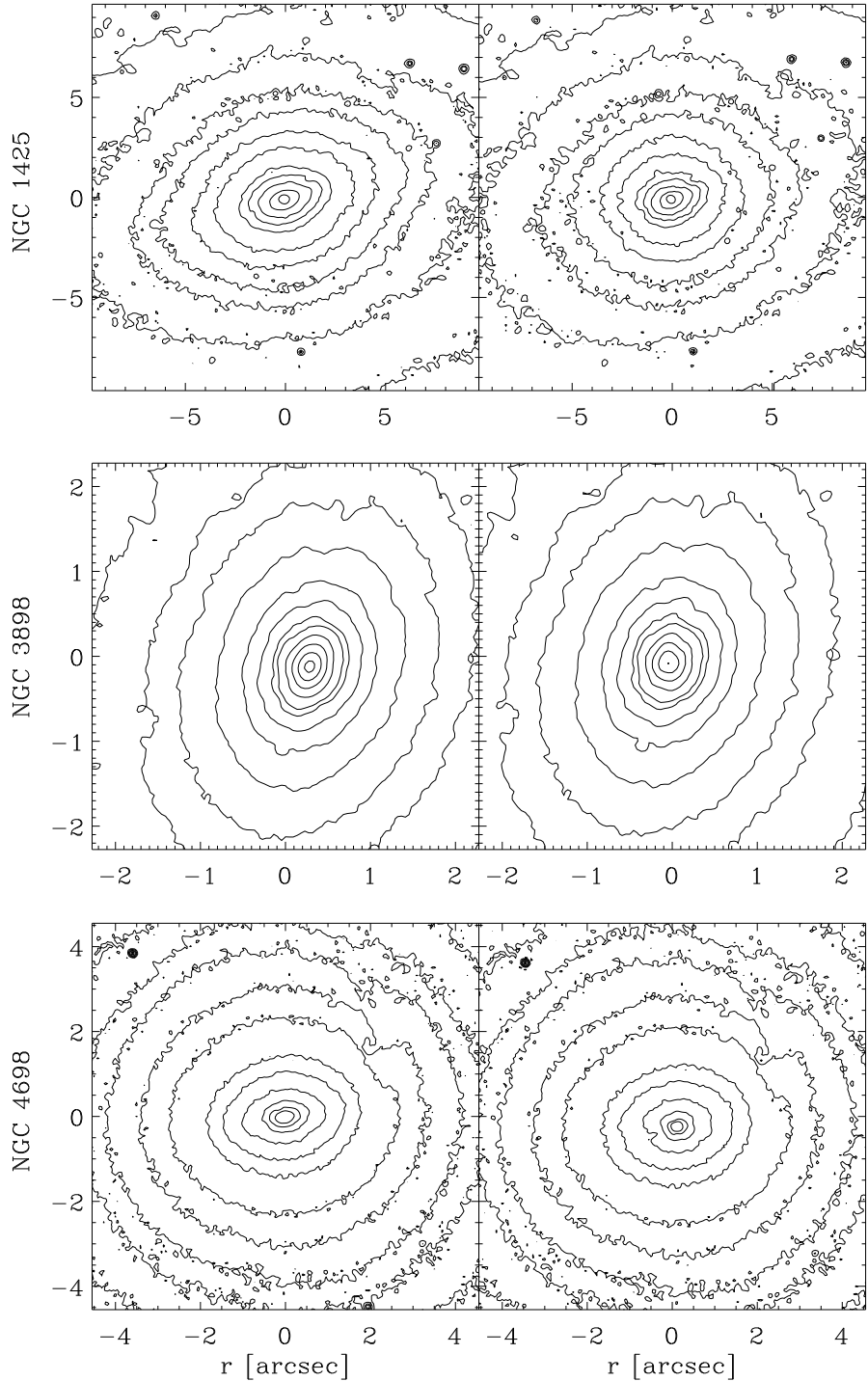} 
\figcaption{Contour plots of the WFPC2/F606W deconvolved images of NGC 
  1425, NGC 3898, and NGC 4698 before ({\it left panel\/}) and after 
  ({\it right panel\/}) the nuclear disk subtraction. Scales are 
  in arcsec and orientations are the same as in Fig. 1.} 
\end{figure} 

\begin{figure} 
\epsscale{0.8} 
\plotone{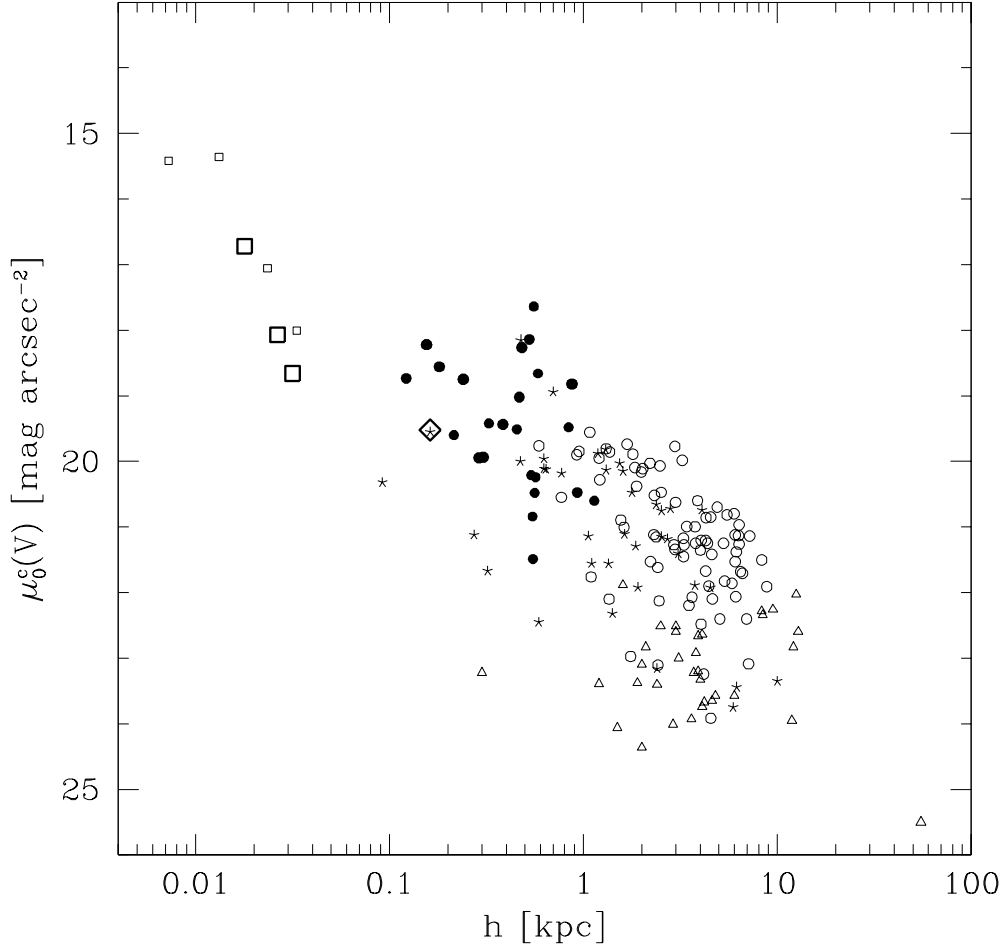} 
\figcaption{ 
  $\mu_0^c$--$h$ diagram adapted from van den Bosch (1998).
  {\it Open circles\/} refer to HSB spirals, {\it triangles\/} to LSB 
  spirals, {\it stars\/} to S0's, {\it filled circles\/} to disky 
  ellipticals, and {\it small squares\/} to nuclear disks in NGC 
  3115, NGC 4342, NGC 4570 and NGC 4594. {\it Large squares\/} 
  correspond to the nuclear disks found in this paper, while the {\it large diamond \/}
 indicates the second larger disk found in NGC 1425 (see Tab. 1 and Sec. 3). The central 
inclination-corrected surface brightness of nuclear disks 
is $\mu_0^c\,=\,\mu_0\,-2.5\,\log({\cos{i}})$.}  
\end{figure} 
 
\clearpage

\begin{deluxetable}{cllccccccccc}
\tablecaption{Parameters of the nuclear disks and host galaxies.}
\tablehead{
 \multicolumn{7}{c}{host galaxy} & & \multicolumn{4}{c}{nuclear disk}\\
\cline{1-7} \cline{9-12}
\multicolumn{1}{c}{Name} & 
\multicolumn{2}{c}{Type} &  
\multicolumn{1}{c}{$D$} & 
\multicolumn{1}{c}{$M^0_{B_T}$} & 
\multicolumn{1}{c}{$i_{gal}$}&  
\multicolumn{1}{c}{PA$_{gal}$} & 
\multicolumn{1}{c}{} &
\multicolumn{1}{c}{$\mu_0$} &  
\multicolumn{1}{c}{$h$} &  
\multicolumn{1}{c}{$i$} & 
\multicolumn{1}{c}{PA}   \\ 
\multicolumn{1}{c}{NGC} &  
\multicolumn{1}{c}{[RC3]} & 
\multicolumn{1}{c}{[RSA]} & 
\multicolumn{1}{c}{[Mpc]} &  
\multicolumn{1}{c}{[mag]} & 
\multicolumn{1}{c}{[$^\circ$]}& 
\multicolumn{1}{c}{[$^\circ$]}&
\multicolumn{1}{c}{} &
\multicolumn{1}{c}{[mag/$''^{2}$]} & 
\multicolumn{1}{c}{[pc]} &  
\multicolumn{1}{c}{[$^\circ$]} &
\multicolumn{1}{c}{[$^\circ$]} }
\startdata
1425   & Sb(s) & Sb(r) & 13.1 & $-19.29$  & 63       & 129     & & 16.90 &  26 & 70 & 137\\ 
\nodata&\nodata&\nodata&\nodata&\nodata   & \nodata  & \nodata & & 18.53 & 162 & 66 & 135\\
3898   & Sab(s)& Sa    & 16.4 & $-19.48$  & 54       & 107     & & 15.36 &  18 & 73 & 102\\ 
4698   & Sab(s)& Sa    & 12.6 & $-19.04$  & 52       & 170     & & 17.27 &  32 & 74 &  71\\
\enddata
\tablecomments{Distances are from Tully (1988) with $H_0=100$ \kmsmpc. 
Absolute magnitudes, position angles and inclinations of the host
galaxies are derived from RC3.}
\end{deluxetable}

\end{document}